\documentclass[onefignum,onetabnum]{siamart190516}

\usepackage{lipsum}
\usepackage{amsfonts}
\usepackage{bm}
\usepackage{nicefrac}
\usepackage{graphicx}
\usepackage{epstopdf}
\usepackage{algorithmic}
\ifpdf
  \DeclareGraphicsExtensions{.eps,.pdf,.png,.jpg}
\else
  \DeclareGraphicsExtensions{.eps}
\fi

\newsiamremark{remark}{Remark}
\newsiamremark{hypothesis}{Hypothesis}
\crefname{hypothesis}{Hypothesis}{Hypotheses}
\newsiamthm{claim}{Claim}

\headers{Density Fluctuations in Stochastic Kinematic Flows}{J. Worsfold, T. Rogers, and P. Milewski}

\title{Density Fluctuations in Stochastic Kinematic Flows\thanks{Submitted to the editors \today.
\funding{This work was funded by EPSRC}}}

\author{Jeremy R. Worsfold
\and Tim Rogers
\and Paul Milewski\thanks{Department of Applied Mathematics, The University of Bath, Bath BA2 7AY, UK (\email{jw3286@bath.ac.uk}, \email{ma3tcr@bath.ac.uk}, \email{pam28@bath.ac.uk}).}}

\usepackage{amsopn}

\ifpdf
\hypersetup{
  pdftitle={Density Fluctuations in Stochastic Kinematic Flows},
  pdfauthor={J. Worsfold, T. Rogers, and P. Milewski}
}

\usepackage[nolist,nohyperlinks]{acronym}
\acrodef{pde}[PDE]{partial differential equation}
\acrodefplural{pde}[PDE]{partial differential equations}
\acrodef{sde}[SDE]{stochastic differential equation}
\acrodefplural{sde}[SDE]{stochastic differential equations}
\acrodef{spde}[SPDE]{stochastic partial differential equation}
\acrodefplural{spde}[SPDE]{stochastic partial differential equations}
\acrodef{spide}[SPIDE]{stochastic partial integro-differential equation}
\acrodefplural{spide}[SPIDE]{stochastic partial intego-differential equations}
\acrodef{lwr}[LWR]{Lighthill Whitham Richards}

\renewcommand{\vec}[1]{\bm{{#1}}}
\renewcommand{\d}{\;\mathrm{d}}
\newcommand{\fs}[1]{\hat{#1}}
\newcommand{\convolve}[2]{\left({#1} \ast {#2}\right)}
\newcommand{\pderiv}[3][]{\frac{\partial^{#1}#2}{\partial #3^{#1}}}
\newcommand{\deriv}[3][]{\frac{\mathrm{d}^{#1}#2}{\mathrm{d}#3^{#1}}}

\DeclareMathOperator{\diag}{diag}

\begin{document}

\maketitle

\begin{abstract}
    At the macroscopic scale, many important models of collective motion fall into the class of kinematic flows for which both velocity and diffusion terms depend only on particle density. When total particle numbers are fixed and finite, simulations of corresponding microscopic dynamics exhibit stochastic effects which can induce a variety of interesting behaviours not present in the large system limit. In this article we undertake a systematic examination of finite-size fluctuations in a general class of particle models whose statistics correspond to those of stochastic kinematic flows. Doing so, we are able to characterise phenomena including: quasi-jams in models of traffic flow; stochastic pattern formation amongst spatially-coupled oscillators; anomalous bulk sub-diffusion in porous media; and travelling wave fluctuations in a model of bacterial swarming.
\end{abstract}



\section{Introduction}
\label{section:intro}

Kinematic flows are systems in which the velocity field of particles in a volume of space can be expressed entirely in terms of their density. As well as continuous media, these flows have been employed as \textit{ad hoc} descriptions of the movement or alignment of large populations of individuals such as in the modelling of pedestrian and animal motion. The other dominant modelling paradigm in these fields are stochastic microscopic (sometimes \textit{individual-based} or \textit{agent-based}) dynamics. Making a precise connection between these is not immediate, and furthermore there are important questions over the size and effect of fluctuations induced by the discreteness of particles which are not present in deterministic, continuum descriptions. These issues have been studied extensively since McKean's seminal paper in 1967 \cite{mckean} on stochastic particles interacting with each other and the limiting distribution as the number of particles tends to infinity. While much of the work done focuses on this limiting distribution \cite{chaintron2021propagation}, fluctuations for finite particle numbers have also been examined \cite{hitsuda1986tightness,Sznitman1984,sznitman1986propagation}.

Mean-field interacting particle models, such as the one proposed by McKean, are applicable to a wide variety of areas of research because of the natural way in which they model particle-like objects interacting with each other. Most notably, the stochastic Kuramoto model \cite{strogatz1991,acebron2005kuramoto} is a specific case of this general paradigm and is a key model in the study of synchronisation. In biology, interacting particle systems have been used to model movement of insects such as the dispersion grasshoppers \cite{aikman1972experimental,lucon2015large}. Models for dispersal and aggregation of organisms have been proposed \cite{grunbaum1994modelling,grunbaum1994translating,milewski2008,Mogilner99anon-local,okubo2001diffusion} at both the individual particle level and continuum limit. Dean (1996) \cite{Dean_1996} not only found a \ac{spide} for the density but expressed the system in terms of its free energy, allowing for connections to be made to real particle systems in physics. Specific \acp{pde} from physics arise from this model such as the porous media equation \cite{blanchard2010probabilistic} and the granular media equation \cite{godinho2015propagation,bolley2013uniform}.

The majority of the previous work mentioned above focuses on particles with Fickian (independent) noise. This is not always the case, however, as it has been observed that organisms can change their diffusivity based on their local density \cite{lucon2015large}. It is therefore useful to generalise the work done by Dean to allow for non-linear diffusions. By doing so, a wide range of swarming and dispersion models can be re-interpreted in terms their individual constituents. 

Here, we examine a class of models obtained as a slight specialization of the model proposed by McKean, in which particle interactions are based on pair-wise distances. By introducing the concept of density of particles as Dirac measures, any integrable function can be fully expressed in terms of this empirical density.  Furthermore, this system is the same in law to the solution of a certain SPIDE containing a spatiotemporal noise term of order $\mathcal{O}(N^{-\nicefrac{1}{2}})$ which can have profound effects for low numbers of particles or systems near criticality. To quantify these effects, we explore the statistics of first-order corrections as generalized Ornstein-Uhlenbeck (OU) processes in both stationary and non-stationary regimes.

This paper is structured as follows. First, we state our model for the behaviour of the individual particles and recast it as an effective SPIDE for the empirical density. Beginning with an exploration of fluctuations around uniform density states, we derive the general expression for the power spectral density. We apply this to a simple model for traffic corresponding in the macroscopic limit to the Lighthill Whitham Richards (LWR) model \cite{Whitham-Lighthill}; there, fluctuations manifest as travelling density waves with a characteristic velocity moving backwards relative to the flow of traffic. Demonstrating the generalization of our technique to higher dimensions, we show how to characterize stochastic patterning the developed in a spatially extended Kuramoto model. In the later sections we explore fluctuations around non-stationary states including models of porous media and biological aggregation. Our methods provide a general framework for the study of noise-induced phenomena in a broad class of models relevant to collective motion.

\section{Kinematic Interacting Particle Systems}

\label{section:model}
We consider $N$ particles in a compact one-dimensional domain $A$, with boundary conditions to be specified later. Direct particle interactions are pairwise, additive, and depend only on the displacement between particles, not their absolute position in the domain. Each particle also experiences a stochastic positional noise, with the strength of the noise also given as an additive function of pairwise particle distance. The position, $X_i$, of particle $i=1,\dots,N$ evolves according to the stochastic differential equation (SDE) 
\begin{equation}
    \d X_i (t) = \frac{1}{N}\sum_{j=1}^{N}f(X_i(t)-X_j(t))\d t + \sqrt{2D}\left[\frac{1}{N}\sum_{j=1}^{N}g(X_i(t)-X_j(t))\right]^{\beta}\d{W_i}(t),
\label{eq:generalmodel}
\end{equation}
where $D,\beta\in\mathbb{R}_+$, $W_n(t)$ is a standard one-dimensional Wiener process independent of other particles. Here, $f$ and $g$ are generalised functions which respectively describe the contributions to the drift and diffusion of particle $i$ from pairwise interactions with other particles. The parameter $\beta$ controls the scaling of the diffusive term, with typical choices for most applications being $\beta = 1$ or $\beta = 1/2$.

Our main object of interest is the particle density, defined as a weighted sum of Dirac masses:
\begin{align}
    \varrho(x,t) = \frac{1}{N}\sum_{n=1}^{N}\delta_{X_n(t)}(x).
\end{align}
The number of particles within some region $A'\subseteq A$ at time $t$ is  $N\int_{A'}\varrho(x,t)\d x$. More generally, if $F$ is an arbitrary function, we have
\begin{align}
    \langle F,\varrho\rangle =\int_A F(x) \varrho(x,t)\d x= \frac{1}{N}\sum_{n=1}^{N}F(X_n(t))\,.
\label{eq:weakform}
\end{align}
Note that $\langle F,\varrho\rangle$ is a random variable; it is the empirical average of $F$ over the (random) empirical density $\varrho$. Fluctuations in particle locations will induce corresponding fluctuations in the measurement $\langle F,\varrho\rangle$, and making judicious choices of $F$ will allow us to characterize various interesting phenomena in these models.
Rather than work directly with particle locations, we find it more convenient to consider a statistically equivalent \ac{spide} formulation. Under mild smoothness assumptions one can show (see \cref{appendix:stochpde} and \cite{Sznitman1984}) that $\langle F,\varrho\rangle$ has the same distribution as $\langle F,\rho\rangle$, where $\rho(x,t)$ satisfies 
\begin{equation}
    \pderiv{\rho}{t} = \pderiv{}{x}\Big(\convolve{f}{\rho}\rho\Big) + D\pderiv[2]{}{x}\Big(\convolve{g}{\rho}^{2\beta}\rho\Big) + \sqrt{\frac{2D}{N}}\pderiv{}{x}\Big(\sqrt{\rho}\convolve{g}{\rho}^{\beta}\eta\Big)\,.
\label{eq:generalspde}
\end{equation}
Here $\eta$ is spatiotemporal white noise, and $\ast$ denotes the convolution $(a\ast b)(x) = \int_A a(x-y) b(y) \d y$ for some functions $a,b:A\mapsto \mathbb{R}$ which also obey appropriate boundary conditions. Note that, crucially, the equivalence in law $\langle F,\varrho\rangle \sim \langle F,\rho\rangle$ is exact for any $N$.

In the limit of large $N$, it is possible to establish a law of large numbers result $\varrho\to\rho^\infty$ solving the \ac{pde}
\begin{equation}
    \pderiv{\rho^\infty}{t} = \pderiv{}{x}\Big(\convolve{f}{\rho^\infty}\rho^\infty\Big) + D\pderiv[2]{}{x}\Big(\convolve{g}{\rho^\infty}^{2\beta}\rho^\infty\Big)\,.
\label{eq:generalpde}
\end{equation}
A corresponding central limit theorem exists quantifying fluctuations around $\rho^\infty$ for large but finite $N$ \cite{Sznitman1984}.

In what follows we will explore the range of behaviours possible in kinematic interacting particle systems of the type specified here. In particular, we will investigate the role of correlations in fluctuations around the \ac{pde} limit in shaping emergent noise-driven dynamics.

\section{Fluctuations around a Uniform Density State}
\label{section:flucsuniformstate}

In many cases a uniform distribution of particles is stable for some or all choices of the model parameters. In this section we study the behaviour of the fluctuations about this state in the case that $\beta=1$. Despite the system being in this stable state, these fluctuations can have visible structure akin to stochastic Turing patterns \cite{mckane2014stochastic,stochasticTuringBrusselator}. These patterns have been observed in cases such as the stochastic Kuramoto model \cite{lucon2011}. If we choose $F=e^{ikx}/2\pi,k\in\mathbb{Z}$ and impose periodic boundary conditions on $A=[-\pi,\pi]$, then $\fs{\rho}_k:=\langle e^{ikx}/2\pi,\rho\rangle$ are the Fourier modes of the density. We can then write \cref{eq:generalspde} with $\beta=1$ in its equivalent Fourier series form as
\begin{equation}
    \deriv{\fs{\rho}_k}{t} = A_k(\vec{\fs{\rho}}) + \frac{1}{\sqrt{N}}\sum_{n=1}^N G_{kn}(\vec{\fs{\rho}})\eta_n
\label{eq:generalfouriersde}
\end{equation}
with
\begin{subequations}
\begin{align}
    A_k(\vec{\fs{\rho}}) & = - 2\pi ik\sum_{\ell\in\mathbb{Z}} \fs{f}_{\ell}\fs{\rho}_{\ell}\fs{\rho}_{\ell-k} - 4\pi^2k^2D \sum_{\ell,q\in\mathbb{Z}} \fs{g}_{\ell}\fs{g}_q\fs{\rho}_{\ell}\fs{\rho}_{q}\fs{\rho}_{k-\ell-q} 
\end{align}
\end{subequations}
and where the noise term has a correlation structure defined by
\begin{align}
    B_{k\ell}(\vec{\fs{\rho}}) := \sum_{n=1}^N G_{kn}(\vec{\fs{\rho}})(G^\dagger(\vec{\fs{\rho}}))_{n\ell} = 4\pi Dk\ell \sum_{j,q\in\mathbb{Z}}\fs{g}_j\fs{g}_{q}^\dagger\fs{\rho}_j\fs{\rho}_q\fs{\rho}_{k-\ell-q-j}
\end{align}
where $G^\dagger$ is the Hermitian conjugate of $G$. This can be derived by using It\^o calculus on the Fourier series expansion of the density
\begin{align}
    \fs{\rho}_k(t) = \frac{1}{2\pi N}\sum_{n=1}^Ne^{-ikX_n(t)}
\end{align}
which can be found in \cref{appendix:stochpde}.

We assume \cref{eq:generalpde} admits a stable, uniform density state, $\rho^*(x,t)=1/2\pi$, and that the fluctuations in each Fourier mode, $k$, can be expressed as
\begin{align}
    \xi_k(t) = \sqrt{N}\left(\fs{\rho}_k(t)-\fs{\rho}^*_k(t)\right)
    \label{eq:fourierLNA}
\end{align}
where $\fs{\rho}^*_k(t)=\delta_{0k}/2\pi$ and $\delta$ is the Kronecker delta. Expanding \cref{eq:generalfouriersde} using \cref{eq:fourierLNA} and keeping terms up to $\mathcal{O}(N^{-\nicefrac{1}{2}})$ gives
\begin{align*}
    \deriv{\xi_k}{t} = \sum_{\ell\in\mathbb{Z}}J^*_{k\ell}\xi_l(t) + \sum_{n=1}^{N} G^*_{kn}\eta_n(t)
\end{align*}
where $J_{kl}:=\partial A_k/\partial \fs{\rho}_l$ is the Jacobian and the $^*$ indicates they are being evaluated at uniform density. Explicitly, we find the Jacobian and noise correlation in this state to be 
\begin{align*}
    J^*_{kl} = \delta_{kl}\left[-ik(\fs{f}_0+ \fs{f}_{k}) - Dk^2\fs{g}_0\left(\fs{g}_0 + 2\fs{g}_{k}\right)\right], \qquad B^*_{k\ell} = \frac{\delta_{k\ell}k\ell D\fs{g}_0^2}{2\pi^2}.
\end{align*}
Note that the Jacobian recovers the linear stability condition that if $J^*_{kk}<0,\forall k$ then the uniform state is stable in the deterministic limit $N\rightarrow\infty$. Assuming this is the case, each Fourier mode is damped according to the eigenvalues of the Jacobian while the noise term continually excites each mode. Depending on the structure of $J^*$ and $G^*$, this can result in spatiotemporal patterns. To quantify this we take the Fourier transform in time. This produces a linear system of decoupled Fourier modes
\begin{equation*}
    \sum_{\ell\in\mathbb{Z}}(-i\omega\delta_{k\ell} - J_{k\ell}^*)\tilde{\xi}_\ell(\omega) = \sum_{n=1}^N G_{kn}^*\tilde{\eta}_n(\omega)
\end{equation*}
with $\tilde{\xi}_k(\omega)$ representing the Fourier transform of $\xi_k(t)$ with frequency $\omega$ and where $\langle\tilde{\eta}_n(\omega)\tilde{\eta}_m(\omega')\rangle=\delta_{nm}\delta(\omega-\omega')$. For convenience we define, $\Phi_{k}(\omega):=-i\omega-J^*_{kk}$, and thus the fluctuations can be expressed as
\begin{align}
    \tilde{\xi}_k(\omega) = \Phi_{k}^{-1}(\omega) \sum_{n=1}^NG_{kn}^*\tilde{\eta}_n(\omega).
    \label{eq:homogeneousflucs}
\end{align}
Since these fluctuations are a stationary, random process we use the Wiener-Khinchin theorem \cite{wiener1930generalized} to represent this process in terms of its spectral decomposition. The power spectrum is given by the average magnitude of \cref{eq:homogeneousflucs},
\begin{align}
    \langle|\tilde{\xi}_k(\omega)|^2\rangle & = \frac{1}{|\Phi_{k}(\omega)|^2}\sum_{n,m=1}^N G^*_{kn}(G^{*\dagger})_{mk}\langle\tilde{\eta}_n\tilde{\eta}_m\rangle  = \frac{B^*_{kk}}{|\Phi_{k}(\omega)|^2}
\label{eq:powerspectrum}
\end{align}
In general, the peak of the power spectrum at $\tilde{\xi}_{k_{\text{max}}}(\omega_{\text{max}})$ will give the characteristic length scale of the fluctuations, $2\pi/k_{\text{max}}$, and their periodicity $2\pi/\omega_{\text{max}}$. These patterns can be stationary or oscillatory, which we now illustrate with two examples.

\subsection{Traffic modelling}
\label{section:traffic}
Often, individual-based modelling of vehicle traffic is done using discrete cellular automata or so-called follow the leader models \cite{helbing2001traffic,followtheleader}. These can be effective at recreating features of traffic flow but are difficult to connect to continuum models. These continuum models are also vital in vehicle traffic modelling since they allow for fast simulations of large road networks through numerical \ac{pde} solutions while still capturing many features of traffic \cite{helbing2001traffic}. Models based on McKean's mean-field interacting diffusion model \cite{chaintron2021propagation} have not been used in the context of vehicle traffic despite them being readily expressed as continuum flows in the limit $N\rightarrow\infty$ and naturally incorporating stochasticity on the individual level.

It has been noted that the viscous Burgers equation arises in the case of linear diffusion ($g(x)=1$) and purely local coupling in the drift choosing the drift term to be a Dirac delta, $f(x)=\delta(x)$ \cite{bonilla1998exactly,bossy1997burgers}. While the \ac{lwr} model is able to capture the propagation of traffic jams as kinematic waves, it cannot explain the formation of phantom jams from uniform flow. By keeping a linear diffusion ($g=1$) but choosing $f(x)=v_0(1-\delta(x)/\rho_{\text{jam}})$ we obtain the \ac{lwr} model with a diffusion term. Explicitly, the density equation is
\begin{align*}
    \pderiv{\rho}{t} + \pderiv{}{x}\left(\rho u\right) = D \pderiv[2]{\rho}{x}
\end{align*}
with $u(\rho)=v_0(1-\rho/\rho_{\text{jam}})$. This diffusion term is commonly used in finite difference schemes for the original \ac{lwr} model \cite{Daganzo1995FDE} to prevent the formation of shock fronts. On an individual level this can be understood as the particles (cars) trying to move at the desired maximum speed, $v_0$, but moving at a reduced velocity which is dependent on the local density. The linear noise term reflects the random, imperfect decisions the particles make, irrespective of the particles around them. 

\begin{figure}[t]
    \centering
    \includegraphics[width=0.95\linewidth]{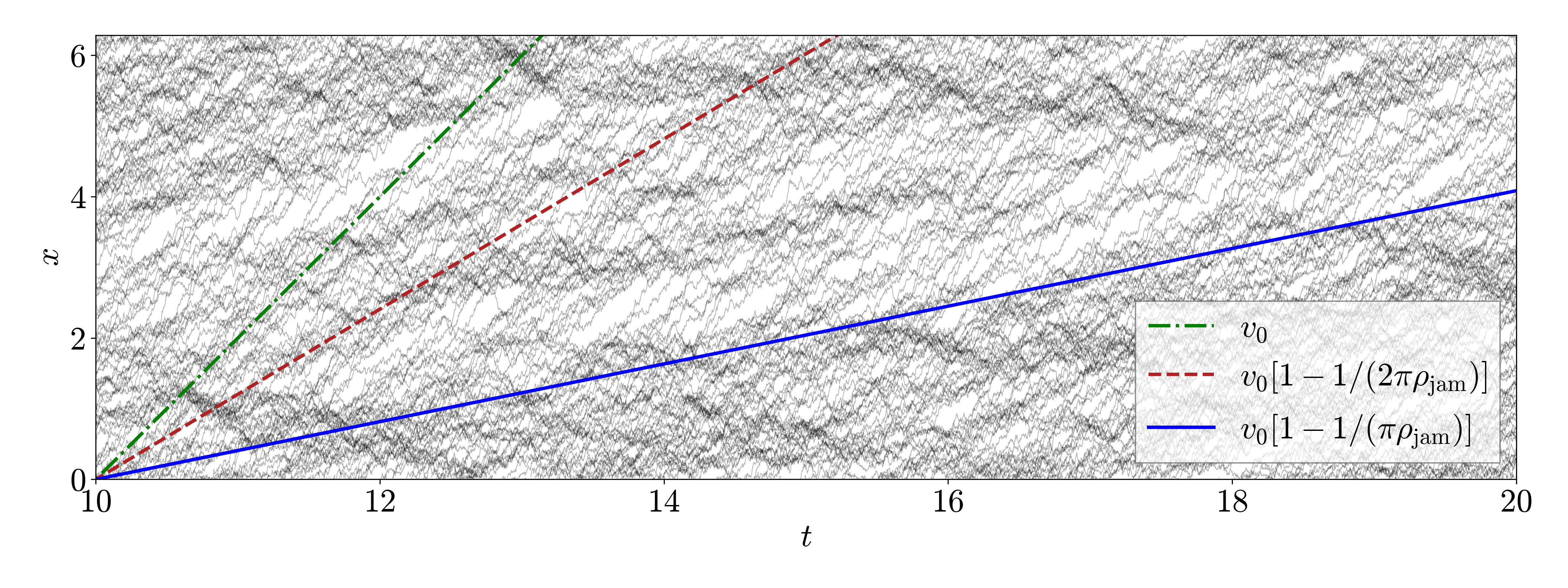}
    \caption{Quasi-Phantom Jams moving at the characteristic velocity, $v_c=v_0(1-2\rho^*/\rho_{\text{jam}})$ with $\rho_{\text{jam}}=0.2,\; v_0=0.5$ and $D=1$. Grey lines show the trajectories in position, $x$, against time, $t$, of $N=200$ particles obeying the stochastic LWR model and forming density clusters matching those predicted in \cref{eq:powerLWR}. Also shown for comparison are the theoretical values for the characteristic wave velocity (solid blue); the average speed of the particles in uniform density (dashed red); and the maximum velocity (dash-dotted green). }
    \label{fig:whithamflucs}
\end{figure}

The uniform density state is always stable for this model, irrespective of the choices of $\rho_{\text{jam}}$,$v_0$ and $D$. Using \cref{eq:powerspectrum}, we expect the fluctuations about this state to obey
\begin{align}
    \langle|\hat{\xi}_k(\omega)|^2\rangle = \frac{k^2D}{2\pi^2\left[(kv_0(1-2\rho^*/\rho_{\text{jam}})-\omega)^2+k^4D^2\right]}.
\label{eq:powerLWR}
\end{align}
Therefore, this model has density waves on the order of $\mathcal{O}(N^{-\nicefrac{1}{2}})$. The speed of these waves, $v_c$, is determined by the peak of the fluctuation spectra, $v_c=\omega_{\text{max}}/ k_{\text{max}}=v_0(1-2\rho^*/\rho_{\text{jam}})$. This is the characteristic speed at $\rho^*$ for the inviscid case, $D=0$. while the average flow of particles is $v_0(1-\rho^*/\rho_{\text{jam}})$. These waves travel backwards relative to the flow of traffic and appear spontaneously, as is the case with phantom jams. In \cref{fig:whithamflucs} we can see these density fluctuations moving backwards relative to the average flow of the particles as predicted.

\subsection{Spatially coupled Kuramoto model}
\begin{figure}[t]
    \centering
    \includegraphics[width=0.98\linewidth]{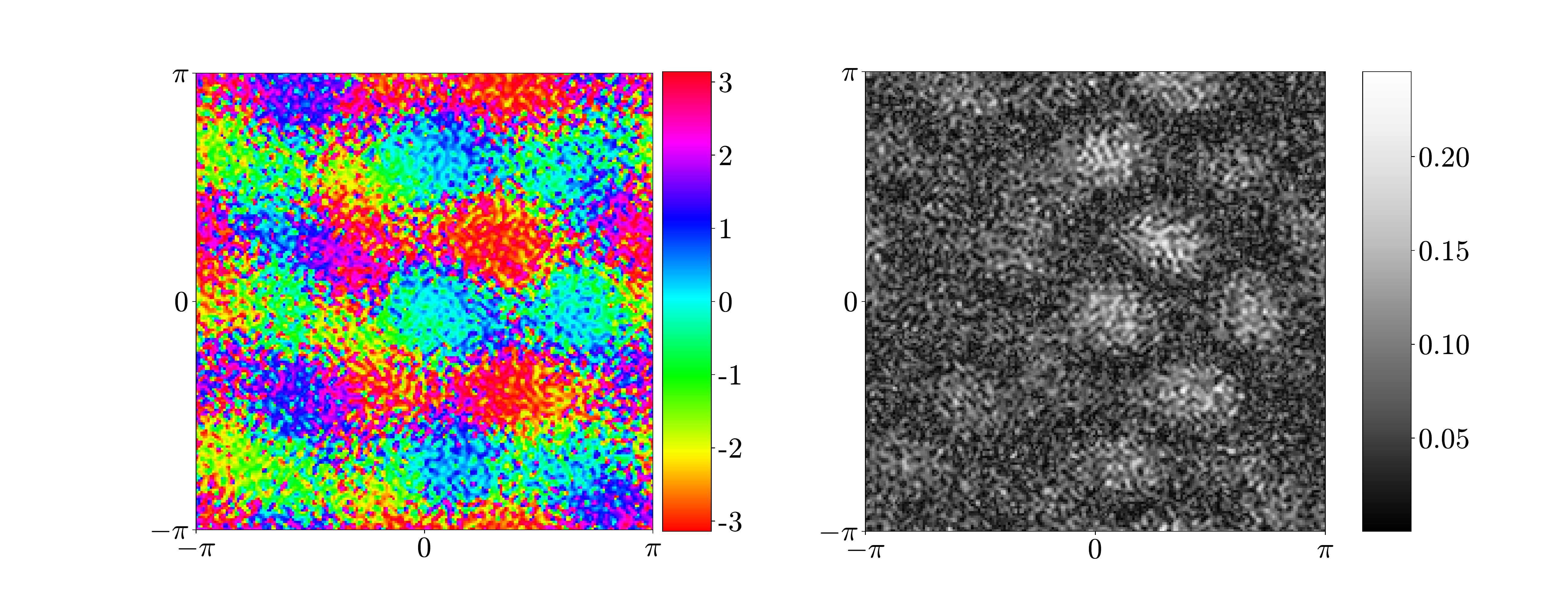}
    \caption{Stochastic patterns for the spatially coupled Kuramoto model. A realisation of the phase, $\psi(\vec{x})$, (left) and coherence, $r(\vec{x})$, (right) for the order parameter $\varphi_1(\vec{x})=re^{i\psi}$, sampled from the zero-mean, Ornstein-Ulhenbeck process in \eqref{eq:spatialkuramotolinearised}.  The spatial coupling is given in \cref{eq:spatialcoupling} with $\kappa=2.6602$, $\sigma_1=0.1$, $\sigma_2=0.05$, $D=0.1$ and $N=10^6$. Clear patterns of local coherence can be seen on the predicted lengthscale of $\mu \approx 3.06$.}
    \label{fig:spatialkuramoto}
\end{figure}

In the case that $f(x)=\sin(x)$ and $g(x)=1$ in \cref{eq:generalmodel}, we obtain the stochastic Kuramoto model \cite{Kuramoto1975InIS,acebron2005kuramoto} with no intrinsic frequencies where $X_n\in[0,2\pi),n=1,\ldots,N$ are then the phases of $N$ coupled oscillators. The Kuramoto model is a simple model for synchronisation but is applicable to many synchronisation phenomena in the real world \cite{MillenniumBridge,Ermentrout1991}. The fluctuations about uniform density for this globally coupled system have been quantified by \cite{lucon2011}. Here we extend the Kuramoto model so that the oscillators have phase, $\vartheta$, and fixed position, $\vec{x}\in [0,L]^d$. Previously, we considered particles with spatial coupling but here the spatial position of the oscillators is fixed and only the phase is variable. The coupling is now a combination of spatial coupling, $K:[0,L]^d\to \mathbb{R}$, and the original phase coupling, $\sin(\vartheta)$. Our modified SDE is thus
\begin{align}
    \d{\vartheta_n} = \frac{1}{N}\sum_{m=1}^N K(\vec{x}_n-\vec{x}_m)f(\vartheta_n-\vartheta_m)\d t + \sqrt{2D}\d{W_n}.
\label{eq:spatialkuramoto}
\end{align}
In this case, we want to study the position-phase density $\rho(\vec{x},\vartheta)=\sum_{n}\delta(\vec{x}-\vec{x}_n)\delta(\vartheta-\vartheta_n)/N$. The analysis follows similarly to before. We express \cref{eq:spatialcoupling} in terms of the Fourier modes of the position-phase density,
\begin{align}
    \d{\fs{\rho}_{\bm{k},\ell}} & = \left[-\ell^2 D\fs{\rho}_{\bm{k},\ell} - 2\pi i \ell L^d \sum_{\bm{k'},\ell'} f_{\ell'} \fs{K}_{\bm{k'}} \fs{\rho}_{\bm{k}-\bm{k'},\ell-\ell'}  \fs{\rho}_{\bm{k'},\ell'}\right]\d t + \frac{1}{\sqrt{N}}\sum_{n=1}^N G_{\bm{k},\ell,n}\d{W_n},
\label{eq:spatialkuramotorho}
\end{align}
where the first index in $\fs{\rho}_{\bm{k},\ell}$ corresponds to the two-dimensional Fourier series in space; the second index is the Fourier series of the oscillator phase; and where G obeys
\begin{align}
    \left[GG^{\dagger}\right]_{(\bm{k},\ell),(\bm{k'},\ell')} = \frac{D\ell\ell'}{\pi L^{d}}\fs{\rho}_{\bm{k}-\bm{k'},\ell-\ell'}.
\end{align}
The phases of the oscillators are uniformly distributed, $\fs{\rho}_{\bm{k},\ell}^\ast = \left(2\pi L^d\right)^{-1}\delta_{\bm{k}\bm{0}}\delta_{\ell0}$, with a small perturbation about this stable state, $\varphi_{\bm{k},\ell} = \sqrt{N}\left(\fs{\rho}_{\bm{k},\ell} - \fs{\rho}_{\bm{k},\ell}^\ast\right)$. As before, use this to linearise \cref{eq:spatialkuramotorho} and keep terms up to $\mathcal{O}(N^{-\nicefrac{1}{2}})$ to obtain
\begin{align}
    \d{\varphi_{\bm{k},\ell}} = \lambda_{\bm{k},\ell}\varphi_{\bm{k},\ell} \d t + \sqrt{\frac{D}{2}}\frac{\ell}{\pi L^d}\d{Z_{\bm{k},\ell}}
    \label{eq:spatialkuramotolinearised}
\end{align}
where $Z_{\bm{k},\ell}$ is a standard complex Wiener process and 
\begin{align}
    \lambda_{\bm{k},\ell} = -\ell^2 D - i\ell\left(f_{\ell} \hat{K}_{\bm{k}} - f_{0} \hat{K}_{\bm{0}}\right).
\end{align}
If we now specify that $f(x)=\sin(x)$, $g(x)=1$ as in the Kuramoto model, then we need only consider the $\ell=1$ Fourier mode, $\varphi_1(\bm{x},t)$. This is a spatially varying version of the complex order parameter often used to describe the Kuramoto model. At each position $\vec{x}$, the magnitude of the order parameter gives the level of coherence of the oscillators while its argument represents the average phase at that location. Given the coupling functions, we have that $\lambda_{\bm{k},1} = -D + \hat{K}_{\bm{k}}$ and thus the system is stable when $D>\fs{K}_{\bm{k}}$. When this is the case, the noise term will still result in spatial structure of order $\mathcal{O}(N^{-\nicefrac{1}{2}})$ for $\varphi_1(\bm{x})$. The strength and lengthscale for this structure is determined by the spatial coupling $K$. As an example, we take the spatial coupling to be 
\begin{align}
    K(\bm{x}) = \frac{\kappa}{\sqrt{2\pi}}\left(\frac{1}{\sigma_1}\exp\left(-\frac{|\bm{x}|^2}{2\sigma_1^2}\right) - \frac{1}{\sigma_2}\exp\left(-\frac{|\bm{x}|^2}{2\sigma_2^2}\right)\right).
    \label{eq:spatialcoupling}
\end{align}
The lengthscale, $\mu=2\pi L/|\vec{k}_{\mathrm{max}}|$, is determined by the closest fourier modes to the maxima of
\begin{align}
    \fs{K}(\bm{k}) = \frac{\kappa}{2\pi L^d} \left[ \exp\left(-\frac{2\pi^2\sigma_2^2|\bm{k}|^2}{L^d}\right) -  \exp\left(-\frac{2\pi^2\sigma_1^2|\bm{k}|^2}{L^d}\right)\right],
\end{align}
the fourier transform of \cref{eq:spatialcoupling}. The maxima are located at 
\begin{align}
    |\bm{k}_{\text{max}}|^2 = \frac{1}{2\pi L^d\left(\sigma_1^2-\sigma_2^2\right)}\ln\left(\frac{\sigma_1}{\sigma_2}\right)
\end{align}
Given the set of equations for the Fourier modes in \cref{eq:spatialkuramotolinearised}, these can be sampled assuming the parameter choices are such that the system is linearly stable. In \cref{fig:spatialkuramoto}, patterns in both the phase and the coherence can be seen showing that certain regions of oscillators develop non-local synchronisation. Over time, these patterns fade and reappear at different positions but always with the same separation $\mu$. Despite the number of particles being large, the parameters are such that the system is close to the point of instability, and so we observe exaggerated patterns relating to strong coherence in the oscillators. Assuming the system behaved deterministically in the large limit would not have resulted in such patterns.

\section{Fluctuations around Non-uniform Density States}
\label{section:heterogeneous}

As in \Cref{section:flucsuniformstate}, we study the fluctuations of order $\mathcal{O}(N^{-\nicefrac{1}{2}})$ in the density about a deterministic solution. Previously, we considered uniform densities but here we look at fluctuations about non-uniform, steady solutions with general $\beta>0$. In general, we have 
\begin{align}
    \rho(x,t) = \rho^*(x,t) + \frac{1}{\sqrt{N}}\xi(x,t)
\label{eq:timedependentLNA}
\end{align}
where $\rho^*(x,t)$ is now a time-dependent solution of \eqref{eq:generalpde}. Expanding \eqref{eq:generalspde} using \eqref{eq:timedependentLNA} and keeping terms up to $\mathcal{O}(N^{-\nicefrac{1}{2}})$ lead us to 
\begin{equation}
\begin{aligned}
    \pderiv{\xi}{t} & = \pderiv{}{x}\left[\xi(f\ast\rho^*) + \rho^*(f\ast\xi)\right] + D\pderiv[2]{}{x}\left[(g\ast\rho^*)^{2\beta}\xi + 2\beta\rho^*(g\ast\rho^*)^{2\beta-1}(g\ast\xi)\right] \\ & \qquad\qquad\qquad + \sqrt{2D}\pderiv{}{x}\left[\sqrt{\rho^*}(g\ast\rho^*)^{\beta}\eta\right].
    \label{eq:linearisedfluctuations}
\end{aligned}
\end{equation}
Non-uniform, time-dependent distributions are likely to arise when considering models for aggregation of organisms such as the compact swarms in \cite{milewski2008,Mogilner99anon-local}.

\subsection{Fluctuations in Porous media}
\label{section:dddiffusion}

In this section we focus on non-linear diffusion of particles by removing the deterministic interactions between particles ($f$ coupling) from \eqref{eq:generalspde}. Also, here we will not be studying travelling wave solutions and thus our domain is $A=(-\infty,\infty)$ meaning we require the density and coupling, $g$, to decay sufficiently at infinity for the convolution to be well defined. The linear noise case where $\convolve{g}{\rho}=1$ is well understood as all particles behave independently and the result is simply Fickian diffusion. 

A general non-linear diffusion is proposed by Okubo and Kareiva \cite{okubo2001diffusion} as an initial dispersal period for a general model for aggregation of insects. Density dependent diffusions have been observed in experiments involving grasshoppers \cite{aikman1972experimental} and are often used to model the group behaviour of organisms. This effect can be replicated by choosing $g(x)=\delta(x)$ in \cref{eq:generalmodel}, meaning that the particles make more pronounced random movements when they are in proximity to other particles in an attempt to find space. Applying this to the large $N$ limit in \cref{eq:generalpde} with $\beta=1$, this becomes the model studied by Okubo and Kareiva:
\begin{align}
    \pderiv{\rho}{t} = \pderiv[2]{}{x}\left(\rho^3\right).
\label{eq:dddiffusion}
\end{align}
Note that here we set $D=1$ without loss of generality as, without a drift, the diffusion coefficient just corresponds to a rescaling of time. For general $\beta$, however, this type of coupling gives rise to a general non-linear diffusion equation:
\begin{align}
    \pderiv{\rho}{t} = \pderiv[2]{}{x}\left(\rho^{1+2\beta}\right) + \frac{1}{\sqrt{N}}\pderiv{}{x}\left(\rho^{\nicefrac{(2\beta+1)}{2}}\eta\right)
\label{eq:stochdiffusionbeta}
\end{align}
In order to quantify the effects of the additional stochastic term, we must first describe the general behaviour of the thermodynamic limit. In the this limit, $N\rightarrow \infty$, \eqref{eq:stochdiffusionbeta} becomes the porous media equation \cite{vazquez2007porous} 
\begin{align}
    \pderiv{\rho^*}{t} = \pderiv[2]{}{x}\left((\rho^*)^m\right), \quad m>1
\label{eq:porousmedia}
\end{align}
with $m=1+2\beta$. Starting from a point source, this equation has a general solution called the Barenblatt solution \cite{Barenblatt} of the form 
\begin{equation}
    \rho^*(x,t) = t^{-a}\left[C - \kappa \left(xt^{-a}\right)^2\right]_+^{\nicefrac{1}{m-1}}, \qquad t>0
    \label{eq:Barenblatt}
\end{equation}
where $a = (m+1)^{-1}$, $\kappa=a(m-1)/ 2m$ and $+$ indicates the positive part of the function ($\rho>0$). For $m=3$, as studied in \cite{okubo2001diffusion}, this takes the form of a semi-ellipse with a growing width, while $m=2$ corresponds to a parabola. The constant $C$ in \eqref{eq:Barenblatt} is determined by assuring the total mass is normalised, $\int_{-\infty}^{\infty} \rho(x,t)\d x = 1$. Thus, we re-express this solution in terms of the height and width of the distribution:
\begin{align}
    \rho^*(x,t) = h(t)\left[1-\left(\frac{x}{r(t)}\right)^2\right]^{\nicefrac{1}{m-1}}.
    \label{eq:heightwidthBarenblatt}
\end{align} 
We determine the height $h(t)$ in terms of the width $r(t)\propto t^{a}$ given that the total mass is unitary (see \Cref{appendix:Barenblatt}). We find that $h(t) = \gamma/r(t)$ with 
\begin{align}
    \gamma = \frac{\Gamma(\frac{m}{m-1}+\frac{1}{2})}{\sqrt{\pi}\Gamma(\frac{m}{m-1})}
\end{align}
and the width of the distribution is given by 
\begin{align}
    r(t) = \gamma^{a(m-1)}\kappa^{-a}t^a.
\end{align}

\begin{figure}
    \centering
    \includegraphics[width=0.95\linewidth]{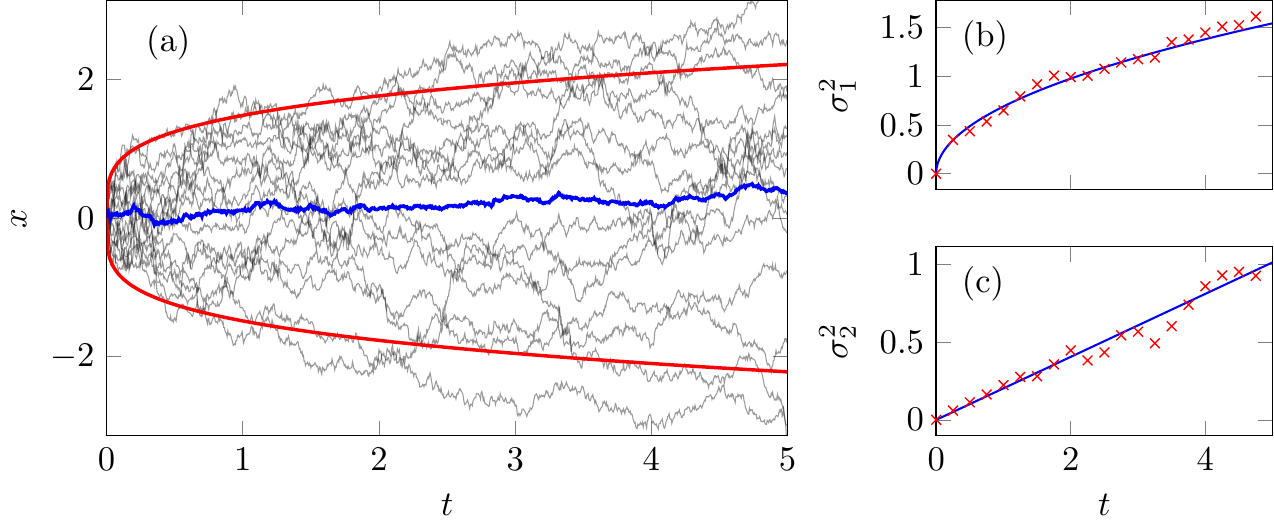}
    \caption{Stochastic porous media equation for $m=3$ ($\beta=1$). (a) Simulation of $N=15$ particles (grey) with the measured centre of mass in blue showing the fluctuations of the first moment, $\Xi_1$, on the order $N^{-\nicefrac{1}{2}}$. The edges of the theoretical distribution are shown in red, illustrating that the particles can extend beyond the theoretical shape. Theoretical ensemble variances over time for the mean displacement (b) and mean squared displacement (c) of particles (blue) compared to the measured variances from 100 simulations of $N=500$ particles (red).}
    \label{fig:ellipseflucs}
\end{figure}

In \Cref{fig:ellipseflucs}(a) it can be seen that for small numbers of particles the shape of the distribution is not well approximated by the solution to the porous media equation and the stochasticity of the particles results in the shape fluctuating over time. To describe this behaviour better, we return to the \ac{spde} of \eqref{eq:stochdiffusionbeta}. In terms of $m$ we have that
\begin{align}
    \pderiv{\rho}{x} = \pderiv[2]{}{x}\left(\rho^{m}\right) + \frac{1}{\sqrt{N}}\pderiv{}{x}\left(\rho^{\nicefrac{m}{2}}\eta\right).
    \label{eq:levylinearisedfluctuations}
\end{align}
The objective is to understand how the shape and position of distribution changes. Since we know its deterministic solution, it is most natural to consider the cumulative moments of the density distribution: $\langle x^n,\rho\rangle,n\in\mathbb{N}_0$. Putting this with \eqref{eq:timedependentLNA}, we define the fluctuations in the cumulative moments as 
\begin{align*}
    \Xi_n(t) & := \sqrt{N}\left(\langle x^n,\rho\rangle - \langle x^n, \rho^*\rangle\right) \\
    & = \int_{-\infty}^{\infty} x^n \xi(x,t) \d x.
\end{align*}
The moments of the distribution are then also fluctuation processes about the deterministic solution. It is well known that finding all the moments of a distribution is equivalent to knowing the full distribution. However, calculating the first few moments will be sufficient in characterising the basic properties of the distribution. Here, we see that the linearised fluctuations in \cref{eq:linearisedfluctuations} becomes
\begin{align}
    \pderiv{\xi}{t} & = m\pderiv[2]{}{x}\big((\rho^*)^{m-1}\xi\big) + \pderiv{}{x}\left((\rho^*)^{\nicefrac{m}{2}}\eta\right).
\end{align}
Substituting in the deterministic solution, $\rho^*(x,t)$, the \ac{spde} for the fluctuations becomes
\begin{align*}
    \pderiv{\xi}{t} & = m h^{m-1}\pderiv[2]{}{x}\left(\left[1-\frac{x^2}{r^2}\right]\xi\right) + \pderiv{}{x}\left((\rho^*)^{\nicefrac{m}{2}}\eta\right) \\
    & = \frac{m h^{m-1}}{r^2}\left[(r^2-x^2)\xi'' - 4x\xi' - 2\xi\right] + \pderiv{}{x}\left((\rho^*)^{\frac{m}{2}}\eta\right)
\end{align*}
where $\xi^\prime=\partial\xi/\partial x$. It is also true that $h^{m-1}r^{-2}=\kappa t^{-1}$ and so
\begin{align}
    \pderiv{\xi}{t} = \frac{m\kappa}{t}\left[(r^2-x^2)\xi'' - 4x\xi' - 2\xi\right] + \pderiv{}{x}\left((\rho^*)^{\nicefrac{m}{2}}\eta\right).
\end{align}
Integrating with respect to the positions weighted by $x^n$ enables us to express the equation above in terms of the cumulative moments 
\begin{align*}
    \int_{-\infty}^{\infty} x^n \pderiv{\xi}{t} \d x & = \frac{m\kappa}{t}\left[ r^2\int_{-\infty}^{\infty} x^n \xi'' \d x - \int_{-\infty}^{\infty} x^{n+2} \xi'' \d x - 4\int_{-\infty}^{\infty} x^{n+1} \xi' \d x \right.\\ & \qquad\qquad \left.- 2\int_{-\infty}^{\infty} x^n \xi \d x\right]  -\int_{-\infty}^{\infty} x^n \left((\rho^*)^{\nicefrac{m}{2}}\eta\right)_x \d x.
\end{align*}
Using integration by parts gives 
\begin{align*}
    \deriv{\Xi_n}{t} & = \frac{m\kappa}{t}\left[n(n-1)r^2 \Xi_{n-2} - (n+2)(n+1)\Xi_n + 4(n+1)\Xi_{n} - 2\Xi_n\right] \\ & \quad\quad  - n\int_{-\infty}^{\infty} x^{n-1} (\rho^*)^{\nicefrac{m}{2}}\eta \d x \\
    & = \frac{mn(n-1)\kappa}{t}\left[r^2 \Xi_{n-2} - \Xi_n\right] - n\int_{-\infty}^{\infty} x^{n-1} (\rho^*)^{\nicefrac{m}{2}}\eta \d x.
\end{align*}
The fluctuations in each moment constitute a set of coupled, time dependent, OU processes which we can express in vector form
\begin{align}
    \dot{\bm{\Xi}}(t) = A(t)\bm{\Xi}(t) + \bm{b}(t)
\label{eq:multivariateOU}
\end{align}
where $\bm{b}$ is a correlated white noise vector with correlation matrix (see \Cref{appendix:porousmediaflucs})
\begin{align*}
    B_{pq}(t) & = 2\int_{-r}^r x^{p+q-2}(\rho^*)^m\d x \\
    & = \begin{cases}
        2pqh^mr^{p+q-1} \frac{\Gamma\left(\frac{p+q-1}{2}\right)\Gamma\left(2+\frac{1}{m-1}\right)}{\Gamma\left(\frac{p+q+3}{2}+\frac{1}{m-1}\right)} & p+q \text{ is even} \\
        0 & \text{otherwise}.
    \end{cases}
\end{align*}
All moments are initially zero since we start from a point source and so the expectation is always zero $\langle\Xi(t)\rangle=0,\forall t>0$. The solution to a multidimensional time dependent OU process has the following covariance \cite{gardiner2009stochastic}
\begin{align*}
\left\langle\Xi(t), \Xi^{\mathrm{T}}(t)\right\rangle = & \exp \left[\int_{0}^{t} A(t') \d t'\right]\left\langle\Xi_0, \Xi_0^{\mathrm{T}}\right\rangle \exp \left[\int_{0}^{t} A^{\mathrm{T}}(t') \d t'\right] \\
& \quad\quad +\int_{0}^{t} \d t' \exp \left[\int_{t'}^{t} A(s) d s\right] B(t') \exp \left[\int_{t'}^{t} A^{\mathrm{T}}(s) d s\right]
\end{align*}
For now we focus on the first element, $B_{11}(t)$, which we calculate to be
\begin{align}
    B_{11}(t) & = 2\gamma^{a(3m-1)}\kappa^{a(m-1)}t^{-a(m-1)} \frac{\Gamma\left(\frac{1}{2}\right) \Gamma\left(2+\frac{1}{m-1}\right)}{\Gamma\left(\frac{5}{2}+\frac{1}{m-1}\right)}.
\end{align}
We also know that $A_{11}(t) = 0$ and so the correlation for the first moment (the centre of mass) is
\begin{subequations}
\begin{align}
    \langle \Xi_1, \Xi_1\rangle & = \int_{0}^t B_{11}(\tau) \d\tau \\
    & = (m+1)\sqrt{\pi}t^{2a}\gamma^{a(3m-1)}\kappa^{a(m-1)}\frac{\Gamma\left(2+\frac{1}{m-1}\right)}{\Gamma\left(\frac{5}{2}+\frac{1}{m-1}\right)}
\end{align}
\end{subequations}
Returning to $m=3$ where the density profile is elliptical, the width grows as $r(t)=2\left(3t/\pi^2\right)^{\nicefrac{1}{4}}$ since $\gamma=2/\pi$. This means that the variance in the centre of mass, $\sigma_1^2=\langle \Xi_1, \Xi_1\rangle$, is 
\begin{align}
    \sigma_1^2 = \frac{\sqrt{3t}}{\pi}.
\end{align}

For the higher moments, this becomes cumbersome to calculate for general $m$ so here we just present the variance in the fluctuations of the mean squared displacement for $m=3$. This time, we have
\begin{align}
    A(t) = \begin{pmatrix} 0 & 0 \\ 0 & -\frac{1}{2t}\end{pmatrix}, \quad B(t) = \begin{pmatrix} \frac{1}{2\pi}\sqrt{\frac{3}{t}} & 0 \\ 0 & \frac{1}{2\pi^2}\end{pmatrix}
\end{align}
and since both matrices are diagonal, the first and second moments are independent. Thus,
\begin{subequations}
\begin{align}
    \sigma^2_2 = \langle \Xi_2, \Xi_2\rangle & = \int_{0}^t B_{22}(\tau)\exp\left[-\int_\tau^t\frac{1}{s}\d s\right]\d\tau  \\
    & = \frac{t}{4\pi^2}.
\end{align}
\end{subequations}
In \Cref{fig:ellipseflucs} we see how the variance of the first two moments increases with time by running an ensemble of 100 simulations compared to these results.

\subsection{Swarming model}
\label{section:swarming}
\begin{figure}[t]
    \centering
    \includegraphics[width=0.8\linewidth]{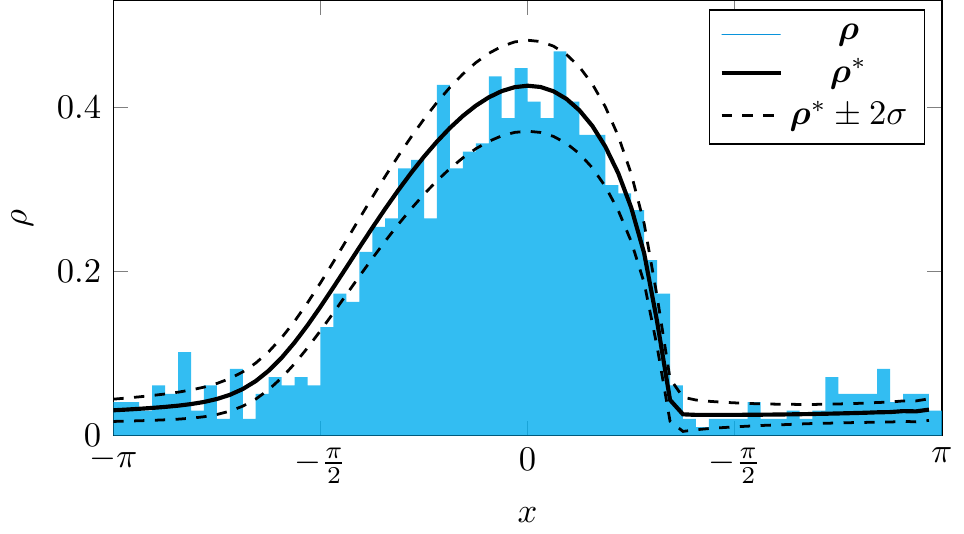}
    \caption{Discretised fluctuations about the deterministic limit for the swarming model proposed by Milewski and Yang \cite{milewski2008} with $M=64$. Solid line: deterministic limit of the swarming model presented in \cref{eq:milewski} for $D=0.4$. Dashed lines: two standard deviations calculated from the diagonal elements of the covariance matrix, $\Sigma$, in \eqref{eq:Lyapunov} from the deterministic limit with $N=1000$. Histogram: distribution of particles from a simulation at a moment in time after the swarm has formed for $N=1000$.}
    \label{fig:swarmingflucs}
\end{figure}

Several models for the aggregation and swarming of organisms in terms of kinematic flows have been proposed \cite{Mogilner99anon-local,milewski2008,topaz2006nonlocal}. The majority of these models can be expressed as the large $N$ limit of our model for appropriate choices of $f,g$ and $\beta$. For example, here we show that the model presented by Milewski and Yang \cite{milewski2008} is recovered if we choose our deterministic coupling to be
\begin{equation*}
f(x) = \begin{cases}
    e^{x}, & x\leq 0 \\
    0, & \mathrm{otherwise}
  \end{cases}
\end{equation*}
and our noise coupling is density-dependent as in the previous section: $g(x)=\delta(x)$ and $\beta=1$. The deterministic coupling is intended to replicate right-sensing organisms attempting to catch up with organisms in front of them and being unaware of those behind them due to their field of view. The resulting PDE from \cref{eq:generalpde} is
\begin{equation}
    \pderiv{\rho}{t}  + \pderiv{}{x}\left(\rho\convolve{f}{\rho}\right) - D\pderiv[2]{}{x}\left(\rho^3\right) = 0.
    \label{eq:milewski}
\end{equation}
Note that this is the same as the kinematic equation in \cite{milewski2008} if $\rho_0=3D/2\pi$. On a periodic domain, this forms a swarm with consistent shape which moves with a particular velocity. The shape is either a continuous distribution of particles with a tail or a compact, more symmetric swarm depending on the choice of $D$. \Cref{fig:swarmingflucs} shows the case where a swarm with a long tail forms. The properties such as the speed and formation of these shapes are studied extensively in \cite{milewski2008}. Qualitatively, the shape of swarm from particle simulations match well with the numerical solutions of the PDE. Going beyond this, we can approximate the fluctuations about the steady state using the numerical solutions and a discretised form of the fluctuations, $\vec{\xi}\in\mathbb{R}^{M}$, with spacing $h = 2\pi/M$. The spatially discretised fluctuations are $\mathcal{O}(\sqrt{M})$ as the fluctuation for a particular bin is related to the number of particles within that bin
\begin{align}
    \vec{\rho}(t) = \vec{\rho}^*(t) + \sqrt{\frac{M}{N}}\vec{\xi}(t).
\end{align}
We introduce $v=\convolve{f}{\rho^*}$ and $\psi=\convolve{g}{\rho^*}$ so \cref{eq:linearisedfluctuations} becomes 
\begin{align}
    \pderiv{\xi}{t} = \pderiv{}{x}(v\xi + \rho^*\convolve{f}{\xi}) + D\pderiv[2]{}{x}\left(\psi^2\xi + 2\rho^*\psi\convolve{g}{\xi}\right) + \sqrt{2D}\pderiv{}{x}(\sqrt{\rho^*}\psi\eta).
    \label{eq:generalflucs}
\end{align}
The discretised form of this equation for the fluctuations is
\begin{align}
    \deriv{\vec{\xi}}{t} = \left[ \Delta (V + \Lambda F) + D \Delta^2 (\Psi^2 + 2\Lambda\Psi G)\right] \vec{\xi} + \Delta \sqrt{\Lambda}\Psi\vec{\eta}.
\end{align}
where $\Delta, \Delta^2$ are the differential and Laplacian finite difference operators respectively and $V,\Lambda,F,\Psi$ are the discretised, matrix counterparts to the coupling and convolution terms in \eqref{eq:generalflucs}. By performing this discretisation, we must treat the stochastic term with care. To obtain the equation above, we first take a grid of points, $i=1,\dots,M$, with equal spacing $h=2\pi/M$ and use a central finite difference scheme, such that 
\begin{align*}
    \Delta_h[\phi_i] &= \frac{1}{2h}[\phi_{i+1}-\phi_{i-1}], \quad
    \Delta^2_h[\phi_i] = \frac{1}{4h^2}[\phi_{i+1}-2\phi_i+\phi_{i-1}].
\end{align*}
The convolutions are approximated as
\begin{align*}
    \convolve{\phi}{a}(x) & = \int_{-\pi}^{\pi} \phi(x-y)a(y)\d y \\
    & \approx h\sum_{j=1}^M \phi_{j-i}a_j
\end{align*}
Hence, for each grid point, $i$, we have that
\begin{align}
    \deriv{\xi_i}{t} & = h\Delta_h\left[v_i\xi_i + \rho_i^*\sum_{j=1}^N f_{j-i}\xi_j\right] + h^2\Delta^2_h\left[\psi_i^2\xi_i + 2\rho_i^*\psi_i\sum_{j=1}^N g_{j-i}\xi_j\right] \\
    & \qquad\qquad +\sqrt{2D}h\Delta_h\left[\sqrt{\rho^*_i}\psi_i\right]\eta_i
\end{align}
where $\langle\eta_i,\eta_j\rangle=\delta_{ij}M=2\pi\delta_{ij}/h$. This dependence on $M$ for scale of the noise is due to the number of particles at each site $i$ being inversely proportional to the number of sites $M$. 

The discretised fluctuations can then be expressed in vector form. Writing $C = \mathrm{circ}(c_1, \dots, c_{M})$ to indicate the circulant matrix where $C_{1j} = c_{j},\;j=1,\dots,N$, the finite difference matrix operators are thus
\begin{align*}
    \Delta & = \mathrm{circ}(0,1,0,\dots,0,-1) \\
    \Delta^2 & = \mathrm{circ}(-2,1,0,\dots,0,1).
\end{align*}
The convolutions can also be written in terms of circulant matrices,
\begin{align*}
    \sum_{j=1}^M \vec{\phi}_{j-i}a_j = \left[\Phi\vec{a}\right]_i
\end{align*}
where $\Phi=\mathrm{circ}(\vec{\phi}_1,\vec{\phi}_2,\dots,\vec{\phi}_M)$. We replace element-wise products of vectors with diagonal matrix-vector products such as $\phi_i a_i=\left[\diag(\vec{\phi})\vec{a}\right]_i$. Using this, we define the following matrices:
\begin{align*}
    \Lambda & = \diag(\vec{\rho^*}), \quad V = \diag(\vec{v}) = \diag(F\vec{\rho^*}), \quad \Psi = \diag(\vec{\psi}) = \diag(G\vec{\rho^*}),
\end{align*}
where $F=\mathrm{circ}(f_1,\dots,f_M)$ and similarly for $G$. Now, we can write the discretised version of \cref{eq:generalflucs} as 
\begin{align}
    \deriv{\vec{\xi}}{t} = \left[ \Delta (V + \Lambda F) + D \Delta^2 (\Psi^2 + 2\Lambda\Psi G)\right] \vec{\xi} + \Delta \sqrt{\Lambda}\Psi\vec{\eta}.
\end{align}
Because $\vec{\xi}$ is a fluctuation about the steady state, it has zero mean and thus we can find its covariance, $\Sigma_{ij}=\langle\xi_i,\xi_j\rangle$, as the solution to the continuous Lyapunov equation:
\begin{align}
    A\Sigma + \Sigma A^T + B = 0
\label{eq:Lyapunov}
\end{align}
where 
\begin{align}
    A = \Delta (V + \Lambda F) + D \Delta^2 (\Psi^2 + 2\Lambda\Psi G)
\end{align}
and
\begin{subequations}
\begin{align}
    B & = \frac{D}{\pi h} \left(\Delta \sqrt{\Lambda}\Psi\right)\left(\Delta\sqrt{\Lambda}\Psi\right)^T \\
    & = \frac{D}{\pi h} \Delta \Lambda\Psi^2 \Delta^T.
\end{align}
\end{subequations}
It is worth reiterating that this holds for general coupling functions $f$ and $g$ for which a numerical approximation to a stationary or travelling wave solution can be obtained. \Cref{fig:swarmingflucs} shows how this can be used for the swarming model mentioned above. The deterministic limit was determined from a numerical simulation by finite differences for a given discretisation. The matrices $A$ and $B$ were calculated using $\vec{\rho}^*$ and solved to give the correlation structure of the bins for a given number of particles in the system. Here, we only show how the diagonal elements of this correlation matrix, $\vec{\sigma}=\sqrt{\diag(\Sigma)}$, illustrate the variation from the deterministic limit.

\section{Discussion}
\label{section:discussion}
The model presented in this paper is both a specification of the model introduced by McKean \cite{Sznitman1984,hitsuda1986tightness} and an extension on the model by Dean \cite{Dean_1996}. The interpretation of the model as a SPIDE for the density allows clear links to kinematic fluid flows to be drawn. Many kinematic flows can be understood as the limit of many interacting particles obeying the simple rules laid out in our model, particularly flows created to model organisms such as fish or insects. Despite focusing on just a few specific examples of kinematic flows, we emphasise that this model encompasses to a large class of kinematic flows. 

By adding a spatial correlation for phase oscillators it was also shown that this model can demonstrate interesting pattern formation for systems close to criticality as the finite-sized effects are amplified. We reiterate that these patterns persist even for large particle numbers close to criticality, a phenomenon not captured by the deterministic limit.

We studied uniform flows for linear diffusion cases and quantified the noise-driven fluctuations. Most notably, a simple individual-based model of traffic related to the LWR model displays quasi-phantom jams due to the finite number of particles. This shows that careful consideration needs to be taken when modelling vehicle traffic as a continuum flow as such effects can be missed. The existence of these quasi-phantom jams warrants further investigation into traffic models of this type. It is possible that extending this to be a second-order model \cite{helbing2001traffic} would allow other features such as actual phantom jams (stop-and-go waves) to be explained. Beyond vehicle traffic, second-order models of this type are used to describe more complex swarming behaviour, for instance those based off the Cucker-Smale model \cite{cuckersmale}. Using the techniques developed here on those models could be of use in the study of other types of swarming behaviour.

This paper examines just a few types of models which can be expressed with specific choices of the coupling functions. In particular, both local and non-local deterministic couplings have been explored with localised or constant noise correlations but non-local noise couplings have not. It is possible such non-trivial noise coupling could result in interesting behaviour.

\bibliographystyle{siamplain}
\bibliography{main}

\begin{thebibliography}{10}

\bibitem{mckean}
{\sc H.~P. McKean}, {\em Propagation of chaos for a class of non-linear
  parabolic equations}, Stochastic Differential Equations (Lecture Series in
  Differential Equations, Session 7, Catholic Univ., 1967),  (1967),
  pp.~41--57.

\bibitem{chaintron2021propagation}
{\sc L.-P. Chaintron and A.~Diez}, {\em Propagation of chaos: a review of
  models, methods and applications}, 2021,
  \url{https://arxiv.org/abs/2106.14812}.

\bibitem{hitsuda1986tightness}
{\sc M.~Hitsuda and I.~Mitoma}, {\em Tightness problem and stochastic evolution
  equation arising from fluctuation phenomena for interacting diffusions},
  Journal of Multivariate Analysis, 19 (1986), pp.~311--328,
  \url{https://doi.org/10.1016/0047-259X(86)90035-7}.

\bibitem{Sznitman1984}
{\sc A.-S. Sznitman}, {\em Nonlinear reflecting diffusion process, and the
  propagation of chaos and fluctuations associated}, Journal of Functional
  Analysis, 56 (1984), pp.~311--336,
  \url{https://doi.org/10.1016/0022-1236(84)90080-6}.

\bibitem{sznitman1986propagation}
{\sc A.-S. Sznitman}, {\em A propagation of chaos result for burgers'
  equation}, Probability theory and related fields, 71 (1986), pp.~581--613,
  \url{https://doi.org/10.1007/BF00699042}.

\bibitem{strogatz1991}
{\sc S.~H. Strogatz and R.~E. Mirollo}, {\em Stability of incoherence in a
  population of coupled oscillators}, Journal of Statistical Physics, 63
  (1991), pp.~613--635, \url{https://doi.org/10.1007/BF01029202}.

\bibitem{acebron2005kuramoto}
{\sc J.~A. Acebr\'on, L.~L. Bonilla, C.~J. P\'erez~Vicente, F.~Ritort, and
  R.~Spigler}, {\em The kuramoto model: A simple paradigm for synchronization
  phenomena}, Rev. Mod. Phys., 77 (2005), pp.~137--185,
  \url{https://doi.org/10.1103/RevModPhys.77.137}.

\bibitem{aikman1972experimental}
{\sc D.~Aikman and G.~Hewitt}, {\em An experimental investigation of the rate
  and form of dispersal in grasshoppers}, Journal of Applied Ecology, 9 (1972),
  pp.~807--817, \url{https://doi.org/10.2307/2401906}.

\bibitem{lucon2015large}
{\sc E.~Lu{\c c}on}, {\em {Large population asymptotics for interacting
  diffusions in a quenched random environment}}, in {Particle Systems and PDEs
  II}, vol.~129 of Springer Proceedings in Mathematics and Statistics, Braga,
  Portugal, Dec 2013, {Springer}, pp.~231--251,
  \url{https://doi.org/10.1007/978-3-319-16637-7}.

\bibitem{grunbaum1994modelling}
{\sc D.~Gr{\"u}nbaum and A.~Okubo}, {\em Modelling social animal aggregations},
  in Frontiers in Mathematical Biology, S.~A. Levin, ed., Berlin, Heidelberg,
  1994, Springer Berlin Heidelberg, pp.~296--325,
  \url{https://doi.org/10.1007/978-3-642-50124-1_18}.

\bibitem{grunbaum1994translating}
{\sc D.~Gr{\"u}nbaum}, {\em Translating stochastic density-dependent individual
  behavior with sensory constraints to an eulerian model of animal swarming},
  Journal of Mathematical Biology, 33 (1994), pp.~139--161,
  \url{https://doi.org/10.1007/BF00160177}.

\bibitem{milewski2008}
{\sc P.~A. Milewski and X.~Yang}, {\em {A simple model for biological
  aggregation with asymmetric sensing}}, Communications in Mathematical
  Sciences, 6 (2008), pp.~397 -- 416, \url{https://doi.org/cms/1214949929}.

\bibitem{Mogilner99anon-local}
{\sc A.~Mogilner and L.~Edelstein-keshet}, {\em A non-local model for a swarm},
  J. Math. Biol, 38 (1999), pp.~534--570,
  \url{https://doi.org/10.1007/s002850050158}.

\bibitem{okubo2001diffusion}
{\sc A.~Okubo and P.~Kareiva}, {\em Some Examples of Animal Diffusion},
  vol.~14, Springer, 2001, \url{https://doi.org/10.1007/978-1-4757-4978-6_6}.

\bibitem{Dean_1996}
{\sc D.~S. Dean}, {\em Langevin equation for the density of a system of
  interacting langevin processes}, Journal of Physics A: Mathematical and
  General, 29 (1996), pp.~L613--L617,
  \url{https://doi.org/10.1088/0305-4470/29/24/001}.

\bibitem{blanchard2010probabilistic}
{\sc P.~Blanchard, M.~Röckner, and F.~Russo}, {\em {Probabilistic
  representation for solutions of an irregular porous media type equation}},
  The Annals of Probability, 38 (2010), pp.~1870 -- 1900,
  \url{https://doi.org/10.1214/10-AOP526}.

\bibitem{godinho2015propagation}
{\sc D.~Godinho and C.~Quiñinao}, {\em {Propagation of chaos for a subcritical
  Keller–Segel model}}, Annales de l'Institut Henri Poincaré, Probabilités
  et Statistiques, 51 (2015), pp.~965 -- 992,
  \url{https://doi.org/10.1214/14-AIHP606}.

\bibitem{bolley2013uniform}
{\sc F.~Bolley, I.~Gentil, and A.~Guillin}, {\em Uniform convergence to
  equilibrium for granular media}, Archive for Rational Mechanics and Analysis,
  208 (2013), pp.~429--445, \url{https://doi.org/10.1007/s00205-012-0599-z}.

\bibitem{Whitham-Lighthill}
{\sc M.~J. Lighthill and G.~B. Whitham}, {\em On kinematic waves ii. a theory
  of traffic flow on long crowded roads}, Proceedings of the Royal Society of
  London. Series A. Mathematical and Physical Sciences, 229 (1955),
  pp.~317--345, \url{https://doi.org/10.1098/rspa.1955.0089}.

\bibitem{mckane2014stochastic}
{\sc A.~J. McKane, T.~Biancalani, and T.~Rogers}, {\em Stochastic pattern
  formation and spontaneous polarisation: The linear noise approximation and
  beyond}, Bulletin of Mathematical Biology, 76 (2014), pp.~895--921,
  \url{https://doi.org/10.1007/s11538-013-9827-4}.

\bibitem{stochasticTuringBrusselator}
{\sc T.~Biancalani, D.~Fanelli, and F.~Di~Patti}, {\em Stochastic turing
  patterns in the brusselator model}, Phys. Rev. E, 81 (2010), p.~046215,
  \url{https://doi.org/10.1103/PhysRevE.81.046215}.

\bibitem{lucon2011}
{\sc E.~Luçon}, {\em Quenched limits and fluctuations of the empirical measure
  for plane rotators in random media.}, Electronic Journal of Probability, 16
  (2011), \url{https://doi.org/10.1214/ejp.v16-874}.

\bibitem{wiener1930generalized}
{\sc N.~Wiener}, {\em Generalized harmonic analysis}, Acta mathematica, 55
  (1930), pp.~117--258.

\bibitem{helbing2001traffic}
{\sc D.~Helbing}, {\em Traffic and related self-driven many-particle systems},
  Rev. Mod. Phys., 73 (2001), pp.~1067--1141,
  \url{https://doi.org/10.1103/RevModPhys.73.1067}.

\bibitem{followtheleader}
{\sc M.~Bando, K.~Hasebe, A.~Nakayama, A.~Shibata, and Y.~Sugiyama}, {\em
  Dynamical model of traffic congestion and numerical simulation}, Phys. Rev.
  E, 51 (1995), pp.~1035--1042, \url{https://doi.org/10.1103/PhysRevE.51.1035}.

\bibitem{bonilla1998exactly}
{\sc L.~L. Bonilla, C.~J. P\'erez~Vicente, F.~Ritort, and J.~Soler}, {\em
  Exactly solvable phase oscillator models with synchronization dynamics},
  Phys. Rev. Lett., 81 (1998), pp.~3643--3646,
  \url{https://doi.org/10.1103/PhysRevLett.81.3643}.

\bibitem{bossy1997burgers}
{\sc M.~Bossy and D.~Talay}, {\em A stochastic particle method for the
  mckean-vlasov and the burgers equation}, Mathematics of Computation, 66
  (1997), pp.~157--192, \url{https://doi.org/10.2307/2153648}.

\bibitem{Daganzo1995FDE}
{\sc C.~F. Daganzo}, {\em A finite difference approximation of the kinematic
  wave model of traffic flow}, Transportation Research Part B: Methodological,
  29 (1995), pp.~261--276,
  \url{https://doi.org/https://doi.org/10.1016/0191-2615(95)00004-W}.

\bibitem{Kuramoto1975InIS}
{\sc Y.~Kuramoto}, {\em Self-entrainment of a population of coupled non-linear
  oscillators}, in International Symposium on Mathematical Problems in
  Theoretical Physics, H.~Araki, ed., Berlin, Heidelberg, 1975, Springer Berlin
  Heidelberg, pp.~420--422, \url{https://doi.org/10.1007/BFb0013365}.

\bibitem{MillenniumBridge}
{\sc B.~Eckhardt, E.~Ott, S.~H. Strogatz, D.~M. Abrams, and A.~McRobie}, {\em
  Modeling walker synchronization on the millennium bridge}, Phys. Rev. E, 75
  (2007), p.~021110, \url{https://doi.org/10.1103/PhysRevE.75.021110}.

\bibitem{Ermentrout1991}
{\sc B.~Ermentrout}, {\em An adaptive model for synchrony in the firefly
  pteroptyx malaccae}, Journal of Mathematical Biology, 29 (1991),
  pp.~571--585, \url{https://doi.org/10.1007/BF00164052}.

\bibitem{vazquez2007porous}
{\sc J.~L. V{\'a}zquez}, {\em The porous medium equation: mathematical theory},
  Oxford University Press on Demand, 2007.

\bibitem{Barenblatt}
{\sc B.~Gilding and L.~Peletier}, {\em On a class of similarity solutions of
  the porous media equation}, Journal of Mathematical Analysis and
  Applications, 55 (1976), pp.~351--364,
  \url{https://doi.org/https://doi.org/10.1016/0022-247X(76)90166-9}.

\bibitem{gardiner2009stochastic}
{\sc C.~Gardiner}, {\em Stochastic methods}, vol.~4, Springer Berlin, 2009.

\bibitem{topaz2006nonlocal}
{\sc C.~M. Topaz, A.~L. Bertozzi, and M.~A. Lewis}, {\em A nonlocal continuum
  model for biological aggregation}, Bulletin of mathematical biology, 68
  (2006), p.~1601, \url{https://doi.org/10.1007/s11538-006-9088-6}.

\bibitem{cuckersmale}
{\sc F.~Cucker and S.~Smale}, {\em Emergent behavior in flocks}, IEEE
  Transactions on Automatic Control, 52 (2007), pp.~852--862,
  \url{https://doi.org/10.1109/TAC.2007.895842}.

\end{thebibliography}

\appendix

\section{Stochastic PDE}
\label{appendix:stochpde}

For any sufficiently smooth function $F$ supported on $A\subseteq \mathbb{R}$, 
\begin{equation}
    \mathbb{E}\left[\frac{1}{N}\sum_{i=1}^NF(X_i) - \int_A \varrho(x,t) F(x) \mathrm{d}x\right] = 0
\label{eq:expectationequivalence}
\end{equation}
where $\varrho$ is the solution to the \ac{spide} in \cref{eq:generalmodel} if $\mathbb{E}\left[\frac{1}{N}\sum_iF(X_i(0)) - \left<\varrho_0,F\right>\right] = 0$.
Firstly, using It\^o calculus we know that
\begin{align*}
    \mathrm{d}F(X_i)&  = \left[F'(X_i)\left(f\ast \rho\right)(X_i) + DF''(X_i)\left(g\ast \rho\right)^{2\beta}(X_i)\right]\mathrm{d}t \\ & \quad\quad + \sqrt{2D}F'(X_i)\left(g\ast \rho\right)^{\beta}(X_i)\mathrm{d}W_i(t)
\end{align*}
and we have used the fact that $\frac{1}{N}\sum_i g(x- X_i)=\int_A\rho(x-y,t) g(y)\mathrm{d}y =: (g\ast\rho)(x,t)$ and similarly for $f$. If we then add the contributions of all particles this becomes
\begin{align*}
    \frac{1}{N}\sum_i\frac{\mathrm{d}}{\mathrm{d}t}F(X_i)&  = \frac{1}{N}\sum_i\left[F'(X_i)\left(f\ast \rho\right)(X_i) + DF''(X_i)\left(g\ast \rho\right)^{2\beta}(X_i)\right] \\ & \quad\quad + \frac{\sqrt{2D}}{N}\sum_iF'(X_i)\left(g\ast \rho\right)^{\beta}(X_i)\eta_i.
\end{align*}
Furthermore, we can express the first two terms integrals against the empirical density
\begin{align*}
    \frac{\mathrm{d}}{\mathrm{d}t}\int_A \mathrm{d}x\rho(x,t) F(x) & = \int_A \mathrm{d}x \rho(x,t) \left\{(f\ast\rho)F'(x)+D\left(g\ast\rho\right)^{2\beta}F''(x)\right\} \\ & \qquad + \frac{\sqrt{2D}}{N}\sum_iF'(X_i)\left(g\ast \rho\right)^{\beta}(X_i)\eta_i.
\end{align*}
We can then put the spatial derivatives onto the density terms by integration by parts so that
\begin{align*}
    \frac{\mathrm{d}}{\mathrm{d}t}\int_A \mathrm{d}x\rho(x,t)F(x) & = \int_A \mathrm{d}x F(x)\left\{\partial_x^2\left(D\left(g\ast\rho\right)^{2\beta}\rho\right)-\partial_x\left((f\ast\rho)\rho\right) \right. \\ & \qquad \left. - \frac{\sqrt{2D}}{N}\sum_i\partial_x\left(\delta(X_i-x)\left(g\ast\rho\right)^{\beta}(x)\eta_i\right)\right\}.
\end{align*}
It remains to be shown that the stochastic term above can be replaced with a Gaussian random field which is independent of the individual particle locations. We follow the approach of Dean (1996) \cite{Dean_1996} by showing that the two formulations have the same correlation function. Firstly we write the noise term in the density evolution as,
\begin{align*}
    \xi(x,t) = - \sum_{i=1}^N \partial_x\left(\delta(X_i-x)\eta_i(g\ast\rho)^{\beta}(x,t)\right)
\end{align*}
which has the following correlation function
\begin{align*}
    \langle\xi(x,t),\xi(y,s)\rangle = 2D\delta(t-s)\partial_x\partial_y\left(\rho(x)(g\ast\rho)^{2\beta}(x)\delta(y-x)\right)
\end{align*}
If we choose $\xi'(x,t)=\partial_x\left(\eta(x,t)\sqrt{\varrho}(x,t)(g\ast\varrho)^{\beta}(x,t)\right)$, where 
\begin{equation*}
    \left\langle\eta(x, t) \eta\left(y, s\right)\right\rangle=\delta\left(t-s\right)\delta(x-y),
\end{equation*}
we regain the our original noise correlation since
\begin{equation*}
\begin{split}
    \langle \xi'(x,t) \xi'(y,s) \rangle & = \left\langle \partial_x\left(\eta(x,t)\sqrt{\rho}(x,t)(g\ast\varrho)^{\beta}(x,t)\right)\partial_y\left(\eta(y,s)\sqrt{\rho}(y,s)(g\ast\varrho)^{\beta}(y,s)\right) \right\rangle\\
    & = \delta(t-s)\partial_x\partial_y\left(\delta(x-y)\varrho(x,t)(g\ast\varrho)^{2\beta}(x,t)\right).
\end{split}
\end{equation*}
Now $\xi'(x,t)$ has no explicit dependency on individual particles, $X_n$, and thus we replace the noise term with this new noise field:
\begin{align*}
    \frac{\mathrm{d}}{\mathrm{d}t}\int_A \mathrm{d}x\rho(x,t)F(x) & = \int_A \mathrm{d}x F(x)\left\{\partial_x^2\left(D\left(g\ast\rho\right)^{2\beta}\rho\right)-\partial_x\left((f\ast\rho)\rho\right) \right. \\ & \qquad\qquad\qquad- \left.\sqrt{\frac{2D}{N}}\partial_x\left(\sqrt{\rho}(g\ast\rho)^{\beta}\eta\right)\right\}.
\end{align*}
Therefore, if $\varrho$ is the solution to \eqref{eq:generalspde} then \eqref{eq:expectationequivalence} holds.

This result is a generalisation of the case of constant noise on each agent in \cite{Dean_1996} and is recovered when $ \convolve{g}{\rho}=1 $. Importantly, the stochastic term is of $\mathcal{O}(1/\sqrt{N})$ and so will be important when the number of agents is not large.

\section{Fourier representation of Density}
When the particles are compactly supported in a region $A$, we can express the density in terms of its Fourier series coefficients. Here we use $A=[-\pi,\pi)$ for simplicity. To find the evolution of the Fourier modes of the density,
\begin{align}
    \fs{\rho}_k(t) = \frac{1}{2\pi N}\sum_{n=1}^N e^{-ikX_n(t)},
\end{align}
we first express the coupling between particles in terms of the Fourier modes of the coupling functions:
\begin{align*}
    \frac{1}{N}\sum_{m=1}^N f(X_n-X_m) & = \int_{-\pi}^\pi \rho(x,t) f(X_n-x) \d{x} \\
    & = \frac{1}{N}\sum_{k,\ell}\int_{-\pi}^\pi \fs{\rho}_k e^{ikx}\fs{f}_\ell e^{i\ell(X_n-x)} \d{x} \\
    & = \frac{2\pi}{N}\sum_{k,\ell} \delta_{k,\ell}\fs{\rho}_k\fs{f}_\ell e^{i\ell X_n} \\
    & = \frac{2\pi}{N}\sum_{k} \fs{\rho}_k\fs{f}_{k} e^{ik X_n}
\end{align*}
and similarly for $g$. We can then rewrite \cref{eq:generalmodel} as 
\begin{align*}
    \d{X_n} = & = \frac{2\pi}{N}\sum_{\ell} \fs{\rho}_\ell\fs{f}_{\ell} e^{i\ell X_n}\d{t} + \frac{2\pi\sqrt{2D}}{N}\sum_{\ell} \fs{\rho}_\ell\fs{g}_{\ell} e^{i\ell X_n}\d{W_n}.
\end{align*}
Next we use Ito's lemma,
\begin{align*}
    \d{\fs{\rho}_k} & = \sum_{n=1}^N\pderiv{\fs{\rho}_k}{X_n}\d{X_n} + \sum_{n,m=1}^N\frac{\partial^2 \fs{\rho}_k}{\partial X_n \partial X_m}\d{X_n}\cdot\d{X_m} \\
\end{align*}
to write the Fourier modes of the density independently of the individual particle positions
\begin{equation}
\begin{split}
    \d{\fs{\rho}_k} & = \frac{-ik}{N}\sum_n e^{-ikX_n}\d{X_n} - \frac{2\pi k^2D}{ N}\sum_{n=1}^N\sum_{\ell,m\in\mathbb{Z}}\fs{g}_{\ell}\fs{g}_m\fs{\rho}_{\ell}\fs{\rho}_{m} e^{-i(k-m-\ell)X_n}\d{t} \\
    & = \left[- 2\pi ik\sum_{\ell\in\mathbb{Z}} \fs{f}_{\ell}\fs{\rho}_{\ell}\fs{\rho}_{\ell-k} - 4\pi^2k^2D \sum_{\ell,m\in\mathbb{Z}} \fs{g}_{\ell}\fs{g}_m\fs{\rho}_{\ell}\fs{\rho}_{m}\fs{\rho}_{k-\ell-m} \right]\d{t} \\ & \quad\quad\quad\quad- \frac{ik\sqrt{2D}}{N}\sum_{n=1}^N\sum_{\ell\in\mathbb{Z}} \fs{g}_{\ell}\fs{\rho}_{\ell}e^{-i(k-\ell)X_n}\d{W_n}. \label{eq:rhosde}
\end{split}
\end{equation}
Now we define
\begin{equation*}
\begin{split}
    A_k(\vec{\fs{\rho}}) & = - 2\pi ik\sum_{\ell\in\mathbb{Z}} \fs{f}_{\ell}\fs{\rho}_{\ell}\fs{\rho}_{k-\ell} - 4\pi^2k^2D \sum_{\ell,m\in\mathbb{Z}} \fs{g}_{\ell}\fs{g}_m\fs{\rho}_{\ell}\fs{\rho}_{m}\fs{\rho}_{k-\ell-m}, \\
    G_{kn}(\vec{\fs{\rho}}) & = -\frac{ik\sqrt{2D}}{\sqrt{N}}\sum_{\ell\in\mathbb{Z}} \fs{g}_{\ell}\fs{\rho}_{\ell}e^{-i(k-\ell)X_n}
\end{split}
\end{equation*}
so that
\begin{equation}
    \deriv{\fs{\rho}_k}{t} = A_k(\vec{\fs{\rho}}) + \frac{1}{\sqrt{N}}\sum_{n=1}^N G_{kn}(\vec{\fs{\rho}})\eta_n(t)
    \label{eq:AGexp}
\end{equation}
where $\eta_n(t)$ are zero mean Gaussian random variables with correlator $\langle\eta_i(t)\eta_j(t')\rangle=\delta_{ij}\delta(t-t')$. The noise correlation matrix in Fourier space is,
\begin{equation*}
\begin{split}
    B_{kl}(\vec{\fs{\rho}})  = \sum_{n=1}^N G_{kn}G^\dagger_{nl} = 4\pi Dkl \sum_{j,m\in\mathbb{Z}}\fs{g}_j\fs{g}_{m}^\dagger\varrho_j\varrho_m\varrho_{k-j-m-l}.
\end{split}
\end{equation*}

\section{Barenblatt Solution to the Porous Media Equation}
\label{appendix:Barenblatt}

Firstly we wish to find the relationship between the height and the width of the deterministic distribution. Since the density. $\rho^*(x)$, is normalised and non-zero for $|x|<r(t)$, we know that
\begin{subequations}
\begin{align}
    1 & = \int_{-r}^r h(t)\left(1-\frac{x^2}{r^2}\right)^{\nicefrac{1}{m-1}}\d{x}
\end{align}
thus the height is given by
\begin{align}
    h(t) & = \left[\int_{-r}^r\left(1-\frac{x^2}{r^2}\right)^{\nicefrac{1}{m-1}}\d{x}\right]^{-1} \\
    & = \frac{1}{r}\left[\int_{-1}^1\left(1-y^2\right)^{\nicefrac{1}{m-1}}\d{y}\right]^{-1} := \frac{\gamma}{r}.
\end{align}
\end{subequations}
Now, carrying out this integration, we find that 
\begin{align}
    \int_{-1}^1\left(1-y^2\right)^{\nicefrac{1}{m-1}}\d{y} = \frac{\sqrt{\pi}\Gamma(\frac{m}{m-1})}{\Gamma(\frac{m}{m-1}+\frac{1}{2})}
\end{align}
and so the explicit form of the normalising constant is 
\begin{align}
    \gamma = \frac{\Gamma(\frac{m}{m-1}+\frac{1}{2})}{\sqrt{\pi}\Gamma(\frac{m}{m-1})}.
\end{align}
In order to relate this to the Barenblatt solution, we return to \cref{eq:Barenblatt},
\begin{subequations}
\begin{align}
    \rho(x,t) & = \left[Ct^{-a(m-1)} - \kappa x^2 t^{-a(m+1)} \right]^{\nicefrac{1}{m-1}} \\
    & = \left[Ct^{-a(m-1)} - \kappa x^2 t^{-1} \right]^{\nicefrac{1}{m-1}}.
\end{align}
\end{subequations}
Similarly for \cref{eq:heightwidthBarenblatt}, we see that 
\begin{align*}
    \rho(x,t) = \left[h^{m-1}-x^2\frac{h^{m-1}}{r^2}\right]^{\nicefrac{1}{m-1}}.
\end{align*}
From these two expressions for the distribution it is clear that
\begin{subequations}
\begin{align*}
    h^{m-1}r^{-2} & = \kappa t^{-1}, \quad h^{m-1} = Ct^{-a(m-1)} \\
    \Rightarrow & h(t) = C^{\frac{1}{(m-1)}}t^{-a}
\end{align*}
\end{subequations}
and thus the width of the distribution is given by 
\begin{subequations} 
\begin{align*}
    r^2 & = \kappa^{-1} C t^{-a(m-1)+1} \\
    r(t) & = \kappa^{-1/2}C^{1/2} t^{a}.
\end{align*}
\end{subequations}
With these results, we can express the constant $C$ in terms of our other constants, $\kappa,\gamma$:
\begin{subequations}
\begin{align*}
    \gamma & = h(t)r(t) = C^{\frac{1}{2}+\frac{1}{m-1}}\kappa^{-\frac{1}{2}} \\
    & = \kappa^{-\frac{1}{2}}C^{\frac{m+1}{2(m-1)}} \\
    \Rightarrow \quad C & = \left(\gamma^2\kappa\right)^{a(m-1)}
\end{align*}
\end{subequations}
Consequently, the width, $r(t)$, can also now be expressed solely in terms of time and the parameters determined by $m$,
\begin{subequations}
\begin{align*}
    r(t)  & = \left(\gamma\kappa^{\frac{1}{2}}\right)^{a(m-1)}\kappa^{-\frac{1}{2}}t^a \\
    & = \gamma^{a(m-1)}\kappa^{-\frac{1}{2}\left[\frac{m-1}{m+1}-1\right]}t^a \\
    & =  \gamma^{a(m-1)}\kappa^{-a}t^a.
\end{align*}
\end{subequations}

\section{Porous Media Fluctuations}
\label{appendix:porousmediaflucs}
To find the variance in the centre of mass of the distribution, we first calculate the correlation between any two given moments of the distribution
\begin{align*}
    B_{pq}(t) & = 2\int_{-r}^r x^{p+q-2}(\rho^*)^m\d{x} \\
    & = 2pqh^m\int_{-r}^r x^{p+q-2}\left[1-\left(\frac{x}{r}\right)^2\right]^{\frac{m}{m-1}}\d{x} \\ 
    & = 2pqh^mr^{p+q-1}\int_{-1}^1 y^{p+q-2}\left[1-y^2\right]^{\frac{m}{m-1}}\d{y} \\
    & = \begin{cases}
        2pqh^mr^{p+q-1} \frac{\Gamma\left(\frac{p+q-1}{2}\right)\Gamma\left(2+\frac{1}{m-1}\right)}{\Gamma\left(\frac{p+q+3}{2}+\frac{1}{m-1}\right)} & p+q \text{ is even} \\
        0 & \text{otherwise}
        \end{cases}
\end{align*}
We can also simplify this further by noticing that
\begin{align*}
    h^m r & = \gamma^{m}r^{-(m-1)} = \gamma^{m-a(m-1)^2}\kappa^{a(m-1)}t^{-a(m-1)} \\
    & = \gamma^{a(3m-1)}\kappa^{a(m-1)}t^{-a(m-1)}.
\end{align*}
This means we can write the variance of the fluctuations in the centre of mass as
\begin{align*}
    \langle \Xi_1, \Xi_1\rangle & = \int_{0}^t \d{t'}B_{11}(t') \\
    &  = 2\int_{0}^t \d{\tau}\tau^{a(1-m)}\gamma^{a(3m-1)}\kappa^{a(m-1)}\frac{\Gamma\left(\frac{1}{2}\right) \Gamma\left(2+\frac{1}{m-1}\right)}{\Gamma\left(\frac{5}{2}+\frac{1}{m-1}\right)} \\
    & = t^{2a}\frac{\sqrt{\pi}}{a}\gamma^{a(3m-1)}\kappa^{a(m-1)}\frac{\Gamma\left(2+\frac{1}{m-1}\right)}{\Gamma\left(\frac{5}{2}+\frac{1}{m-1}\right)} \\
    \langle \Xi_1, \Xi_1\rangle & = (m+1)\sqrt{\pi}t^{2a}\gamma^{a(3m-1)}\kappa^{a(m-1)}\frac{\Gamma\left(2+\frac{1}{m-1}\right)}{\Gamma\left(\frac{5}{2}+\frac{1}{m-1}\right)}.
\end{align*}
\end{document}